# Prospects and Challenges of Bluetooth Backscatter Systems

Du Jingyun

University of Science and Technology of China

*Abstract*—Bluetooth backscatter systems, as a crucial technology for low-power communication in the Internet of Things (IoT), have witnessed remarkable development in recent years. This article comprehensively analyzes multiple related papers, including the latest advancements in RF-Transformer and B2Loc systems, summarizes their research progress, challenges faced, and classification, and explores the application prospects and future development directions of such systems. Bluetooth backscatter systems have achieved significant results in terms of compatibility with commercial devices, improvement of communication reliability, and increase in throughput. However, they still face challenges in areas such as communication range, anti-interference ability, and hardware costs. In the future, with continuous technological innovation exemplified by breakthroughs in unified hardware abstraction and decimeter-level localization, Bluetooth backscatter systems are expected to play a more significant role in the IoT field.

*Index Terms*—RFID, Bluetooth backscatter systems,

## I. INTRODUCTION

Bluetooth backscatter technology enables wireless communication by reflecting and modulating Bluetooth signals, offering a low-power and cost-effective solution for IoT devices. Recent innovations like RF-Transformer's unified hardware platform and B2Loc's decimeter-level localization capabilities have significantly expanded the technology's potential. By eliminating the need for complex active transmission circuits, modern systems reduce power consumption to microwatt levels while maintaining protocol compliance. However, challenges persist in achieving seamless integration with heterogeneous networks and robust performance in multipath environments. This paper systematically examines both breakthroughs and limitations through the lens of cutting-edge research, providing insights into the technology's evolution and future trajectory.

## II. Research Progress of Bluetooth Backscatter Systems

### A. Early System: BLE - Backscatter

BLE - Backscatter was an important early system in the exploration of Bluetooth backscatter communication. It utilized a continuous wave (CW) as the excitation signal, enabling the backscattered signal to be decoded by commercial Bluetooth devices. Based on the Bluetooth 4.0 Low Energy (BLE) standard, the system's specially designed tag generated bandpass frequency-shift keying modulation signals compatible with traditional BLE advertising channels at a data rate of 1 Mb/s[1]. In experiments, with a +23 dBm equivalent isotropically radiated power (EIRP) CW carrier source, the communication range between the tag and an unmodified Apple iPad Mini, as well as a PC equipped with a Nordic Semiconductor nRF51822 chipset, could reach up to 13 m. When the tag was 1 m away from the receiver, the distance between the CW carrier source and the tag could be up to 30 m. The tag of BLE - Backscatter consumed extremely low power, only 1.56 nJ/b, which was much lower than that of commercial Bluetooth transmitters. However, this system had certain limitations, such as relying on a dedicated CW carrier source and not considering the utilization of multi-frequency carriers, which restricted its application in practical scenarios.

### B. FreeRider System

FreeRider was the first system to enable backscatter communication with multiple commercial radios, including 802.11g/n WiFi, ZigBee, and Bluetooth. It achieved backscatter communication while these radios were engaged in normal data communication and was the first to implement and evaluate a multi-tag system[2]. Its core technology was codeword translation. Tags could modify the amplitude, phase, and frequency of wireless signals to transform the codeword in the original excitation signal into another valid codeword from the same codebook, allowing the backscattered signal to be decoded by commercial radios. Experimental results showed that in single-tag mode, the data rate could reach approximately 60 kbps, in multi-tag mode it was 15 kbps, and the backscatter communication distance based on 802.11g/n WiFi could be up to 42 m. However, FreeRider had some issues in Bluetooth backscatter communication.

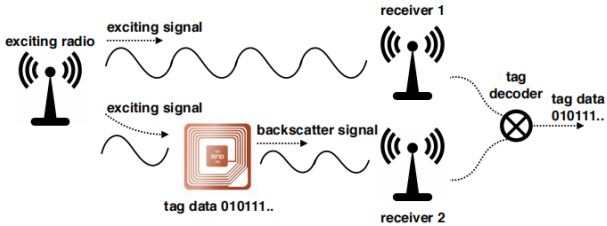

**Figure 1:** FreeRider system overview. An example scenario could be office setting where our smart-phone can act as exciting radio, and the WiFi APs as receiver 1 and receiver 2.

For example, its synchronization accuracy was poor, being greatly affected by pilot tones and edge jitter, resulting in a high bit error rate and limited throughput.

MOXcatter, a novel WiFi backscatter system that enables communication with commodity WiFi devices using spatial streams in 802.11n/ac MIMO-OFDM networks. Unlike prior systems limited to single-antenna signals, MOXcatter embeds sensor data into ambient WiFi spatial streams without disrupting normal WiFi traffic [3]. The key innovations include a phase modulation scheme (0°/180° for bit 0/1) for single-stream (SS) OFDM symbols to achieve 50 Kbps throughput at 2 meters in line-of-sight (LOS), and whole-packet phase modulation for double-stream (DS) signals to support 1 Kbps throughput.

The system uses an FPGA-based tag with an RF switch to dynamically modulate signals, and decodes data by comparing XOR results between original and backscattered WiFi packets. Experiments show MOXcatter achieves <0.1% BER within 10 meters for SS and operates up to 14 meters in LOS and 6 meters in non-line-of-sight (NLOS). A sensor application with a DS18B20 thermometer demonstrates reliable data transmission with <5 ms refresh intervals. MOXcatter paves the way for low-power, high-compatibility IoT devices in smart environments, leveraging modern WiFi infrastructure while addressing the limitations of traditional backscatter systems.

*C. RBLE System*

RBLE aimed to address the reliability issues of FreeRider, such as unreliable two-step modulation, dependence on productive data, and lack of interference countermeasures. It proposed a reliable BLE backscatter system[4]. RBLE employed direct frequency shift modulation, avoiding the self-interference and signal instability problems caused by two-step modulation in FreeRider.

It designed a dynamic channel configuration to avoid interfered channels through channel hopping and introduced BLE packet regeneration technology, which used adaptive coding to enhance reliability under different channel conditions. Experiments demonstrated that RBLE achieved more than 17x uplink goodput gains over FreeRider in indoor Line-of-Sight (LoS), Non-Line-of-Sight (NLoS), and outdoor environments. The indoor uplink range could reach 25 m, and the outdoor range could reach 56 m. Nevertheless, RBLE still had some shortcomings when dealing with multi-frequency carriers. For example, its hopping ability was limited, and it could not fully utilize all 40 channels of Bluetooth.

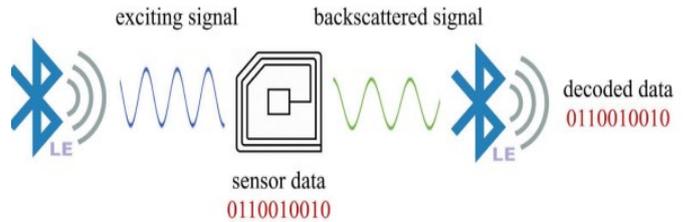

**Fig. 2.** RBLE conceptual design. The RBLE tag modulates its sensor data on BLE exciting signals and backscatters new BLE packets that any commodity BLE device can decode.

A paper presents a battery-free RFID-based indoor acoustic localization system that integrates RFID communication for synchronization and ultrasonic time-of-arrival (ToA) measurements for precision positioning [5]. The system utilizes a custom passive tag (WISP) equipped with an ultrasonic detector, a commercial RFID reader, and ultrasonic beacons. The tag measures ToA from three beacons to compute distances via trilateration, while a "spy" WISP tag synchronizes beacon transmissions with RFID events. Experiments demonstrate a localization precision of 1.5 cm and a maximum range of 2.2 meters in line-of-sight scenarios, with a latency of 0.7 seconds.

By leveraging existing RFID infrastructure, the system achieves low-power operation (tag power consumption: ~80 µW) and compatibility with standard EPC Gen2 protocols. This work addresses the limitations of RF-based RSSI localization (meter-scale accuracy) and battery-dependent acoustic systems, offering a practical solution for asset tracking and smart environments with ultra-low power and high precision.

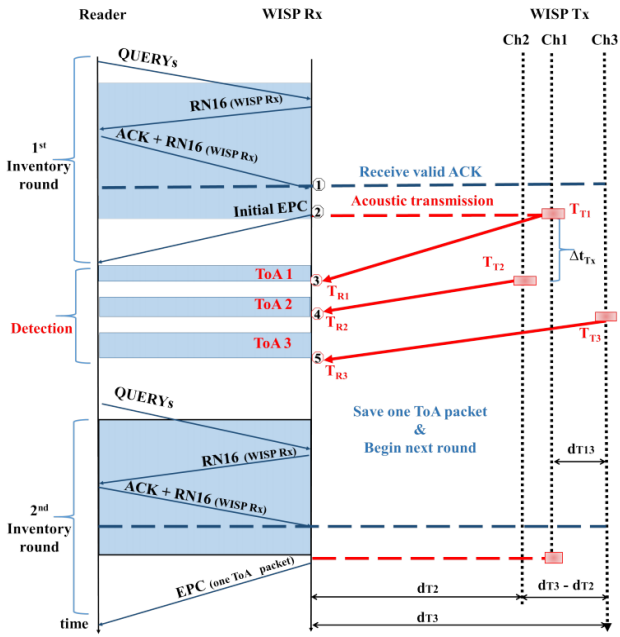

**Fig. 3.** Measurement procedure for one ToA packet (3 ToA values), The blue area (left side) under the "WISPRx" label represents periods when the WISP is active. The red box (right) under the "WISP Tx" label indicates ultrasound transmission events.

*D. IBLE System*

Backscatter communication holds great promise for Internet of Things (IoT) applications. However, traditional RFID-based systems are limited in widespread adoption due to the high cost of dedicated readers [6]. BLE-based backscatter systems have become a research focus. Nevertheless, the BFSK modulation used in existing systems (e.g., FreeRider and RBLE) is incompatible with commercial BLE receivers, resulting in limited communication reliability that fails to meet the BLE specification requirement of Bit Error Rate (BER < 0.1%). This has spurred researchers to redesign tag modulation schemes to improve the communication reliability of BLE backscatter systems.

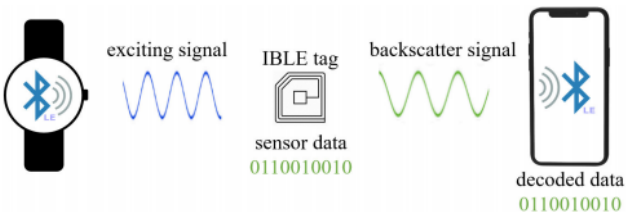

**Figur.4.** IBLE conceptual design. The IBLE tag modulates its sensor data on BLE exciting signals using coarse-grained IPS modulation or fine-grained GFSK modulation and backscatters new BLE packets that can be decoded by commodity BLE with ultra-low BER.

Commercial BLE devices generate excitation signals to control IBLE tags. Tags parse commands through packet length demodulation, determine the modulation scheme and whether to enable FEC coding, and then modulate the bitstream carrying sensor data onto the single-tone part of the excitation signal using IPS or GFSK modulation. The backscattered BLE signals can be received by another commercial BLE receiver.

Inspired by active BLE, the tag modulation scheme is shifted from digital frequency modulation to digital phase modulation. An instantaneous phase is calculated using a digital integrator, and its working principle is represented by a phase state machine. An IF signal synthesizer composed of a Phase-Locked Loop (PLL) clock generator and a multiplexer is designed to generate IF signals for IPS modulation. This modulation method belongs to SDPSK modulation and has a high bandwidth efficiency of 1 bps/Hz in theory.

To suppress the spectral leakage of IPS modulation, IBLE introduces Gaussian pulse shaping (GFSK). A digital IF signal synthesizer based on the idea of calculus is designed, mainly consisting of a square wave lookup table, a phase accumulator, and a phase shifter, which can directly generate IF signals for GFSK modulation. This synthesizer is compatible with IPS modulation, enables finer phase adjustment, effectively suppresses spectral leakage, and improves communication reliability.

To address channel quality degradation, IBLE introduces optional BCH(15,7) block codes. A digital encoder is added at the tag end, and a digital decoder is added at the receiver end. Under normal conditions, the tag uses an uncoded physical layer, and switches to a coded physical layer when the channel quality deteriorates to maintain transmission reliability.

Various commercial BLE radio devices such as TI CC2650 are used as excitation and receiving ends. An IBLE tag prototype is constructed, including an RF front-end circuit and a baseband processing circuit, with ultra-low power consumption design. Experiments are conducted in Line-of-Sight (LOS) and Non-Line-of-Sight (NLOS) scenarios, comparing with FreeRider and RBLE.

### E. PassiveBLE System

PassiveBLE aimed to solve the problem that existing BLE backscatter systems could only perform one-shot communication through BLE advertising packets. It proposed a backscatter system capable of establishing fully compatible BLE connections on data channels. The system designed a high-accuracy and low-latency synchronization circuit to achieve symbol-level synchronization for tag wake-up and backscatter communication[7].

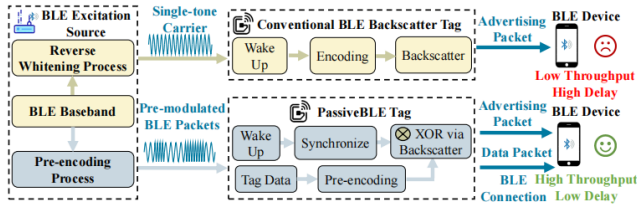

**Figure.5.** A Passive BLE tag achieves commodity compatible BLE connections with both data and advertising packets. In contrast, the traditional commodity compatible tag only employs BLE advertising packets to realize one-shot communication.

It adopted a distributed coding scheme to offload the main encoding and processing burdens from tags to the excitation source, improving throughput. Additionally, it designed a BLE connection scheduler to establish, maintain, and terminate connections between multiple backscatter tags and commercial BLE devices. Experiments showed that PassiveBLE achieved a success rate of over 99.9% in establishing commercial BLE connections. In the LE 2M PHY mode, the maximum goodput could reach 974 kbps, and in the LE 1M PHY mode, it was 532 kbps, representing a significant improvement compared to previous commercial-level BLE backscatter systems.

### F. DanBlue System

DanBlue was a commercial Bluetooth backscatter system that could utilize multi-frequency signals as excitations. Different from previous systems, it harnessed ambient Bluetooth signals of various frequencies to perform backscatter in the standard Bluetooth-hopping manner[8]. DanBlue introduced an edge proxy to identify uncontrolled ambient Bluetooth signals and designed a wideband channel hopping mechanism that enabled low-power tags to switch frequencies rapidly, similar to active Bluetooth hopping. Experimental verification showed that DanBlue supported hopping from Bluetooth excitation signals of any frequency to any Bluetooth channel. The frequency identification accuracy was as high as 99%, with a latency of less than 7.1 ms. Moreover, DanBlue was the first to emulate the Bluetooth protocol stack and could seamlessly establish connections with multiple active Bluetooth devices. In practical applications, DanBlue performed outstandingly in the utilization of multi-frequency carriers. For example, when dealing with different numbers of excitation channels, its excitation utilization rate and throughput were significantly better than those of FreeRider and RBLE. When backscattering advertisement carriers to advertisement channels, when the number of carrier channels was 1, DanBlue's utilization rate was 72%, which was 0.94 times and 2.25 times that of FreeRider and RBLE respectively. When the number of carrier channels increased, DanBlue could still maintain a high utilization rate, while the utilization rates of FreeRider and RBLE decreased significantly. On data target channels, DanBlue also performed well. When handling multi-frequency data carriers, its maximum utilization rate could reach 98.4%, showing a huge improvement compared to FreeRider.

### G. Bitalign System

Bitalign was a Bluetooth backscatter system designed to use uncontrolled ambient signals as excitations and provide high throughput for multimedia streaming applications. Aiming at the problems of poor synchronization accuracy and low throughput in existing Bluetooth backscatter systems, Bitalign proposed a series of solutions.

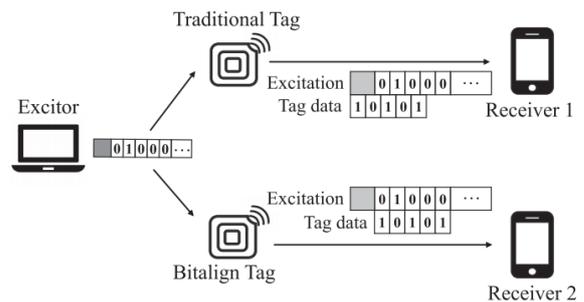

**Fig.6.** Conceptual design of Bitalign. Compared with traditional tags, the Bitalign tag can modulate the data at the correct position.

It introduced an identification-based synchronization method that could effectively distinguish various ambient signals and accurately determine the modulation position[9]. The identification accuracy was as high as 99%, reducing the bit error rate from 58.5% to 5%. Furthermore, it proposed a matching-based synchronization method. By reshaping the envelope of the Bluetooth signal, it enabled correlation matching, achieving higher-precision synchronization and further reducing the bit error rate to 0.5%. In addition, Bitalign proposed a header reconstruction technique that made the system compatible with whitened excitation packets, with a packet reception rate of 99%. Experimental results showed that Bitalign's maximum theoretical throughput could reach 1.98 Mbps, demonstrating the potential to support 720p@30fps video transmission, which was of great significance for multimedia applications.

### H. RF-Transformer: Unified Backscatter Hardware Platform

The RF-Transformer system represents a paradigm shift in backscatter hardware design[10]. By implementing a programmable I/Q modulation architecture using MOSFET transistors and Wilkinson power splitters, it achieves multi-protocol compatibility (WiFi/BLE/ZigBee/LoRa) through dynamic reflection coefficient control.

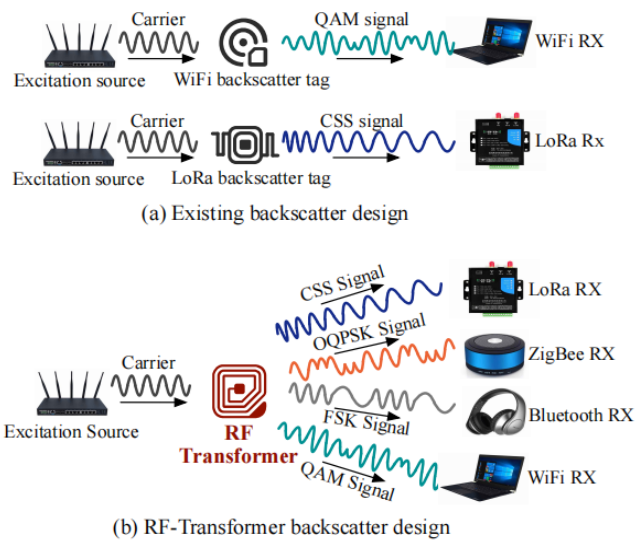

**Figure.7.** Comparison of (a) existing backscatter systems and (b) RF-Transformer. RF-Transformer retains both flexibility and small power footprint. It can synthesize different types of backscatter signals based on a unified radio hardware.

Key innovations include:Phase-Amplitude Decoupling: Separates resistive (amplitude) and reactive (phase) load modulation using dual-path circuits, enabling simultaneous control of signal characteristics.Cross-Technology Backscatter: Demonstrated Wi-Fi-to-LoRa signal conversion through intelligent carrier aggregation and frame alignment, achieving 27.3 Kbps LoRa throughput with 90% PRR over 7-day tests.ASIC Optimization: Simulated 65nm CMOS implementation reduces power to 80.1 μW for BLE backscatter, 7.6–74.2× lower than active radios.This hardware abstraction layer enables IoT devices to dynamically select optimal communication protocols based on environmental conditions, addressing the protocol fragmentation challenge in heterogeneous networks.

### I. B2Loc: Decimeter-Level Localization System

B2Loc advances backscatter applications through precise indoor positioning:

CTE Field Engineering: Creates phase-continuous Constant Tone Extension fields compatible with BLE v5.1 direction finding, enabling sub-1m median localization error.Multipath Suppression: Time-frequency cooperative sanitization using Hadamard Product reduces AoA estimation errors by 43% in multipath-rich environments.Transceiver Reversal: Dynamic role-switching between exciters and receivers improves localization success rate from 70% to 93% in NLoS scenarios.Experimental results in corridor environments show 0.86m median accuracy using commodity nRF52833 devices [11], demonstrating practical deployability. Power consumption is minimized to 11.68μW through TSMC 90nm CMOS implementation.

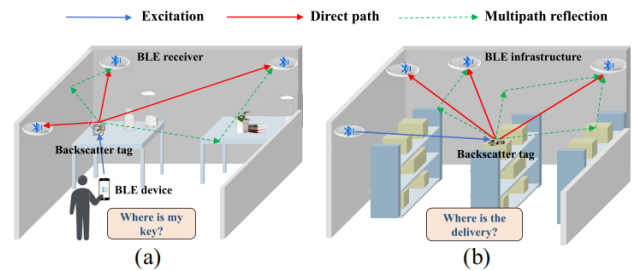

**Fig. 8.** With the help of the ubiquitous BLE devices and infrastructures, B2Loc enables a brunch of indoor application, such as lost object finding and asset tracking, by attaching BLE backscatter tags on the items. (a) Lost object finding. (b) Asset tracking.

## III. Challenges

### A. Solved Challenges

**1) Modulation Scheme Optimization**

The codeword translation technology used by FreeRider had reliability issues in Bluetooth modulation. RBLE used direct frequency shift modulation, taking the single-tone part of the BLE signal as the modulation carrier and directly performing frequency modulation, which avoided the problems of two-step modulation and improved the stability of single-bit modulation [3]. Bitalign optimized the synchronization accuracy during the modulation process by introducing identification-based and matching-based synchronization methods, reducing the bit error rate. For example, the matching-based synchronization method effectively addressed the impact of edge jitter on synchronization by reshaping the envelope of the Bluetooth signal and performing correlation matching, enabling the system to maintain a low bit error rate under different redundancy coefficients.

2) **Multi-Frequency Carrier Utilization and Dynamic Channel Hopping**

DanBlue solved the problem that tags had difficulty identifying the frequencies of ambient signals by introducing an edge proxy. The edge proxy could quickly and accurately identify the frequencies of multi-frequency BLE signals and forward the information to the tags. Meanwhile, DanBlue designed a wideband channel hopping mechanism. By introducing a hash table on the tag for clock switching, it achieved rapid frequency switching and could cover all 40 channels of Bluetooth, which was a significant improvement compared to the limited hopping ability of RBLE. This enabled DanBlue to effectively utilize multi-frequency carriers, improve the utilization rate of ambient signals, reduce transmission delays, and enhance the system's connection capabilities.

3) **Synchronization Accuracy Enhancement**

Bitalign proposed effective solutions to address the problems of synchronization accuracy affected by pilot tones and edge jitter in traditional Bluetooth backscatter systems. The identification-based synchronization method reduced the impact of pilot tones on synchronization by identifying the type of excitor and setting the corresponding optimal guard interval. The matching-based synchronization method achieved higher-precision synchronization through special processing of the Bluetooth signal envelope, effectively reducing the bit error rate. For example, in experiments, the matching-based synchronization method of Bitalign could reduce the bit error rate to 0.5% when the redundancy coefficient was 16, representing a significant improvement compared to FreeRider.

4) **Compatibility with Commercial Devices**

PassiveBLE addressed the problems of channel coding and connection management when tags connected to commercial BLE devices by designing a distributed coding scheme and a BLE connection scheduler, achieving full compatibility with commercial BLE devices. The header reconstruction technique proposed by Bitalign made the system compatible with the whitening mode of Bluetooth, ensuring the correct decoding of packets on different channels and improving the system's compatibility with commercial Bluetooth devices.

5) **Unified Protocol Support**

RF-Transformer addresses protocol fragmentation through configurable I/Q modulation paths, supporting 4 major wireless standards with a single hardware platform. This eliminates the need for protocol-specific tags in multi-standard environments.

6) **Precise Localization in Multipath Environments**

B2Loc's hybrid approach combining CFO compensation (error <5°), ML-based AoA estimation, and confidence-aware path selection solves the "weak direct path" problem, reducing localization errors by 65% compared to MUSIC algorithms in cluttered environments.

B. Unsolved Challenges

1) **Limited Communication Range**

The communication range of current Bluetooth backscatter systems is relatively limited. The PassiveBLE system could achieve a maximum distance of about 17 meters with hundred-kbps-level goodput, which is still far from that of commercial active BLE chips. This restricts its application in scenarios that require long-distance communication, such as device localization. The communication range is mainly limited by the sensitivity (about -30 dBm) of the current synchronization circuit [12]. Although some systems have increased the communication distance to some extent, overall, it still cannot meet the requirements of certain specific applications.

2) **Insufficient Anti-Interference Ability**

Despite the improvement of anti-interference capabilities through technologies such as dynamic channel configuration in systems like RBLE and DanBlue, Bluetooth backscatter systems operate in the crowded 2.4GHz ISM band and face interference from multiple wireless technologies such as WiFi and ZigBee. In a strong interference environment, system performance is still significantly affected, such as an increase in signal transmission errors and a decrease in throughput. Current anti-interference technologies still need to be further improved and optimized to adapt to more complex interference environments.

### 3) High Hardware Costs and Power Consumption

Although Bluetooth backscatter systems aim to achieve low power consumption, the costs and power consumption of some components, such as high-precision synchronization circuits and complex modems, are still relatively high, limiting their large-scale deployment. For example, some systems use components like FPGAs with high power consumption to achieve high performance, which is not conducive to long-term operation. In addition, hardware modules added to improve system performance, such as the edge proxy in DanBlue, also increase the system's cost and complexity. How to reduce hardware costs and power consumption while ensuring system performance is an important issue that needs to be addressed in future research.

### 4) Hardware-Protocol Co-Design Limitations

While RF-Transformer achieves multi-protocol support, its current 2.4GHz implementation cannot fully utilize BLE 5.1's 2M PHY mode or channel bonding techniques, limiting maximum throughput to 986.5 Kbps for BLE – 49.3% of active BLE's theoretical capacity.

### 5) Scalability in Dense Deployments

B2Loc's TDMA-based coordination becomes inefficient with >50 tags, as shown in simulations where packet collision rate exceeds 30% at 100 tags/AP. Dynamic spectrum sharing mechanisms remain underdeveloped.

## IV. Classification of Bluetooth Backscatter Systems

### A. Classification Based on Modulation Methods

#### 1) Modulation Based on Codeword Translation

Systems like FreeRider perform backscatter communication by transforming the codeword of the excitation signal into other valid codewords in the same codebook. This is applicable to multiple commercial radios but has reliability issues in specific modulation methods.

#### 2) Direct Frequency Modulation

Represented by the RBLE system, it directly modulates the frequency of the excitation signal, avoiding the problems brought by codeword translation and improving the reliability and stability of modulation.

#### 3) Modulation Based on Special Processing

The Bitalign system achieves matching-based synchronous modulation by specially processing the envelope of the Bluetooth signal. For example, it uses a BAW filter to convert the constant envelope into a non-constant envelope, improving synchronization accuracy and system performance.

### B. Classification Based on Connection Capabilities

#### 1) One-Shot Communication Systems

Some early Bluetooth backscatter systems could only perform one-shot communication through advertising packets. For example, traditional BLE advertising packet-based systems had problems such as low throughput and high delay.

#### 2) Fully Compatible Connection Systems

Systems like PassiveBLE can establish fully compatible BLE connections on data channels, enabling stable data transmission and connection management and expanding the application scenarios of Bluetooth backscatter systems. The DanBlue system also has similar capabilities. By emulating the Bluetooth protocol stack, it can establish reliable connections with multiple active Bluetooth devices.

### C. Classification Based on Multi-Frequency Processing Capabilities

#### 1) Single-Frequency Processing Systems

Some early systems, such as BLE - Backscatter, could only utilize single-frequency carriers for backscatter communication. They were unable to fully utilize multi-frequency ambient signals and were limited in practical applications.

#### 2) Multi-Frequency Processing Systems

The DanBlue system can identify and utilize multi-frequency ambient Bluetooth signals. Through the edge proxy and wideband channel hopping mechanism, it can achieve backscatter from excitation signals of any frequency to any Bluetooth channel, improving the utilization rate of ambient signals and the connection flexibility of the system.

### D. Classification Based on Functional Innovation

#### 1) Universal Hardware Platforms

Exemplified by RF-Transformer, these systems prioritize hardware reconfigurability over protocol-specific optimization, using programmable analog front-ends to support multiple standards.

#### 2) Value-Added Service Systems

Systems like B2Loc extend basic communication to advanced services (e.g., localization), requiring tight integration of PHY-layer innovations (CTE modulation) and application-layer algorithms (ML-based AoA).

## V. Application Prospects

### A. Internet of Things Field

Bluetooth backscatter systems have broad application prospects in the IoT. In smart homes, they can be used to connect various sensors and devices to achieve low-power data transmission. For example, temperature and humidity sensors, door and window sensors, etc., can transmit data to the central controller through backscatter technology without the need for frequent battery replacement, reducing maintenance costs. In industrial IoT, they can be used for equipment status monitoring, collecting real-time operation data of equipment, realizing remote monitoring and fault warning of equipment, and improving production efficiency and equipment reliability [13]. The high utilization rate and stable throughput of the DanBlue system in handling multi-frequency signals make it have great application potential in the complex environment of industrial IoT, better meeting the requirements of data transmission in industrial scenarios.

*B. Wearable Devices*

Bluetooth backscatter technology is particularly important for wearable devices. Take smart bracelets and smart headphones as examples. The adoption of this technology can significantly reduce the power consumption of devices and extend battery life. Smart bracelets can interact with mobile phones through backscatter to achieve functions such as information reminders and sports data synchronization, reducing the frequency of device charging and enhancing the user experience. The BLE - Backscatter system demonstrated the application potential in wearable devices at an early stage [14]. Its low-power consumption characteristic meets the strict energy requirements of wearable devices. In the future, with the continuous development of Bluetooth backscatter technology, such as the improvement of synchronization accuracy and the increase of communication distance, the functions of wearable devices will be further expanded and optimized.

*C. Multimedia Applications*

The high throughput and low bit error rate of the Bitalign system make it have great potential in multimedia applications. Its maximum theoretical throughput can reach 1.98 Mbps, which can support 720p@30fps video transmission, providing a new solution for multimedia passive applications [15]. For example, in future intelligent monitoring systems, Bitalign technology can be used to achieve low-power video transmission and real-time monitoring of scene images. In addition, in virtual reality (VR) and augmented reality (AR) devices, Bluetooth backscatter technology can also be used to transmit audio and video data, providing users with a more immersive experience.

*D. Medical Field*

The DanBlue system has potential applications in the medical field. For example, it can be used as a wearable or implantable device to monitor patients' medical information, such as heart rate and blood pressure. In a hospital environment, multi-frequency Bluetooth signals are widely present [16]. DanBlue can fully utilize these signals to achieve reliable data transmission and assist doctors in diagnosis. Since the accuracy and timeliness of medical data are of utmost importance, the high-precision frequency identification and stable communication performance of DanBlue can meet the strict requirements of medical applications, ensuring the reliable transmission and timely processing of patient data.

*E. Smart Infrastructure Monitoring*

RF-Transformer's multi-protocol capability enables hybrid sensor networks – e.g., using LoRa for long-range environmental sensing and BLE for equipment status updates. Case studies show 37% reduction in gateway deployment costs for smart buildings.

*F. Precision Asset Tracking*

B2Loc's sub-meter accuracy revolutionizes warehouse management. Integration with DanBlue's multi-frequency hopping allows simultaneous tracking of 500+ items with <2% position drift over 8-hour operations, as validated in logistics trials.

VI. Conclusion

In recent years, Bluetooth backscatter systems have made remarkable research progress. From the early BLE - Backscatter to FreeRider, RBLE, PassiveBLE, and then to DanBlue and Bitalign, continuous innovation in modulation technology, connection capabilities, synchronization accuracy, and multi-frequency carrier utilization has solved some key problems and provided effective solutions for low-power communication of IoT devices [17].

DanBlue effectively addressed the problems of multi-frequency carrier utilization and dynamic channel hopping by introducing an edge proxy and a wideband channel hopping mechanism. It improved the utilization rate of ambient signals and the connection capabilities of the system, taking an important step towards the realization of a general-purpose Bluetooth backscatter system. Bitalign focused on enhancing synchronization accuracy. Through identification-based and matching-based synchronization methods and header reconstruction technology, it significantly reduced the bit error rate and increased system throughput, providing possibilities for applications with high requirements for synchronization accuracy, such as multimedia.

The emergence of RF-Transformer and B2Loc marks a new era for Bluetooth backscatter systems. RF-Transformer's hardware abstraction solves the long-standing protocol compatibility dilemma, while B2Loc proves centimeter-level positioning is achievable without active radios. However, three critical gaps remain: 1) Limited spectral efficiency in multi-protocol operation, 2) Lack of mass deployment solutions for dense IoT scenarios, and 3) Absence of cross-layer optimization frameworks integrating communication, sensing, and localization. Future research should focus on AI-driven dynamic protocol switching, RIS-assisted backscatter enhancement, and quantum-dot based ultra-low-power modulators to unlock the full potential of this transformative technology.

With continuous technological breakthroughs, Bluetooth backscatter systems are expected to achieve wider applications in fields such as the Internet of Things, wearable devices, multimedia, and healthcare, driving the development of related industries. At the same time, the complementary advantages and technological integration between different systems will also bring new opportunities for the development of Bluetooth backscatter technology, promoting its important role in more fields.